\documentclass[useAMS,usenatbib]{mn2e}
\usepackage{graphicx}
\usepackage[usenames,dvipsnames]{xcolor}

\usepackage{epstopdf}

\newcommand{\Line}[3]{\Ion{#1}{#2}\,$\lambda$\,#3}
\newcommand{\Lines}[3]{\Ion{#1}{#2}\,#3}
\newcommand{\Ion}[2]{#1{\,\small #2}}
\newcommand{\Ha}{\mbox{${\mathrm{H\alpha}}$}}

\newcommand{\Msun}{\mbox{$\mathrm{M}_{\odot}$}}

\title[Observations of IR Com in a low state]
{Spectroscopy of the enigmatic short-period cataclysmic variable IR\,Com in an
  extended low state}

\author[C.J. Manser \& B.T. G\"ansicke]{
C.J. Manser\thanks{C.Manser@warwick.ac.uk},
B.T. G\"ansicke\\
Department of Physics, University of Warwick, Coventry CV4 7AL,
UK \\
}

\begin{document}

\date{Accepted 2013. Received 2013; in original form 2013}

\pagerange{\pageref{firstpage}--\pageref{lastpage}} \pubyear{2006}

\maketitle

\label{firstpage}

\begin{abstract}
We report the occurrence of a deep low state in the eclipsing
short-period cataclysmic variable IR\,Com, lasting more than two
years. Spectroscopy obtained in this state shows the system as a
detached white dwarf plus low-mass companion, indicating that
accretion has practically ceased.  The spectral type of the
  companion derived from the SDSS spectrum is M6--7, somewhat later
  than expected for the orbital period of IR\,Com. Its radial
velocity amplitude, $K_2=419.6\pm3.4$\,km/s, together with the
inclination of $75^{\circ}-90^{\circ}$ implies
$0.8\,\Msun<M_\mathrm{wd}<1.0\,\Msun$. We estimate the white dwarf
temperature to be $\simeq15000$\,K, and the absence of Zeeman
splitting in the Balmer lines rules out magnetic fields in excess of
$\simeq5$\,MG. IR\,Com still defies an unambiguous
  classification, in particular the occurrence of a deep, long low
  state is so far unique among short-period CVs that are not strongly
  magnetic.
\end{abstract}

\begin{keywords}
Stars: Dwarf novae -- white dwarfs -- novae, cataclysmic variables --
stars:individual: IR\,Com
\end{keywords}

\section{Introduction}

Cataclysmic Variables (CVs) are binary systems in which a white dwarf
accretes matter from a low mass main sequence star via Roche lobe
overflow (see \citealt{warner95-1} for a comprehensive overview). The
observational appearance of a CV, and therefore its classification
depends only on relatively few fundamental properties, primarily the
mass transfer rate from the companion, the magnetic field of the white
dwarf, the mass ratio, and the orbital period. A small number of CVs
defy a definitive classification, probably straddling the boundary of
one or more of the defining physical properties. Such systems provide
an excellent challenge to test our understanding of the accretion
processes in CVs as well as the evolution of compact binaries
(e.g. \citealt{pattersonetal13-1}).

IR\,Com is an eclipsing CV with an orbital period of 2.09\,h, just at
the lower boundary of the period gap \citep{richteretal97-1}, where
gravitational wave radiation is the dominating angular momentum loss
agent \citep{tutukovetal85-1}. The long-term light curve of
\citet{richteretal97-1} is very atypical for CVs below the period gap,
displaying erratic variations between 16\,mag to 17\,mag, punctuated
very rarely by short bright states reaching 14\,mag, and occasional
fainter states near 18\,mag. As such, it does not resemble the
dominant class of short-period CVs, i.e. dwarf novae which exhibit
quasi-periodic outbursts of their thermally unstable accretion disc
\citep{meyer+meyer-hofmeister81-1}. However, it also does not share
the characteristics of polars, disc-less CVs containing a strongly
magnetic white dwarf, that show irregular long-term brightness
variations related to changes in the mass loss rate of the companion
star (e.g. \citealt{kafka+honeycutt05-1}). Both
\citet{richteretal97-1} and \citet{katoetal02-2} discuss the
possibility of IR\,Com being an intermediate polar (IP), i.e. a CV
with a weakly magnetic white dwarf, but argued against that hypothesis
because of the non-detection of a spin period in the optical
photometry, and of the long duration of one of the well-sampled
outburst-like events. This leaves IR\,Com being one of the few CVs
falling in-between the well-established classes, and strongly suggests
that at least one of its fundamental physical properties must be close
to the limiting value that defines the observational threshold between
the different CV classes.
 
We report the first spectroscopic observation on IR\,Com obtained
during the longest-lasting low state recorded so far, constrain the
stellar parameters, and discuss the possible nature of the system. 

\begin{figure*}
\includegraphics[angle=-90,width=1.3\columnwidth]{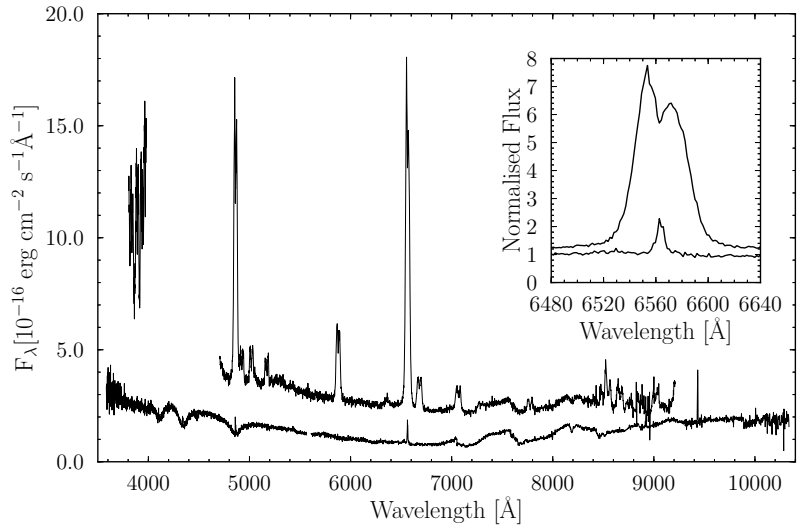}
\caption{\label{f-SDSS} Comparison of the two spectra of IR Com
  obtained by the SDSS in January 2008 (top) and June 2012 (bottom, no
  offset has been applied to the two spectra). The 2008 spectrum is
  typical of a short-period non-magnetic CV, with strong double-peaked
  Balmer and He lines from the accretion disc, and some contribution
  from the companion at $\lambda>7000$\,\AA. In contrast, the 2012
  spectrum is entirely dominated by the white dwarf and its M-dwarf
  companion. The inset shows the \Ha\ emission line normalised to the
  continuum flux, with the 2008 emission line offset by +0.2. For the
  low-state \Ha\ profile, the three sub-spectra obtained in 2012 were
  averaged in the rest frame of the companion star to avoid orbital
  smearing.
\label{f-spectra}}
\end{figure*}

\section{Observations}
Spectroscopy of IR\,Com was obtained by SDSS on two occasions, once
using the original SDSS spectrograph on 15 January 2008
\citep{abazajianetal09-1}, and again on 11 June 2012, this time with
the BOSS spectrograph \citep{ahnetal13-1}. The first spectrum
(Fig.\,\ref{f-spectra}), even though partially corrupted, is typical
of a short-period CV, characterized by a blue continuum superimposed
by strong, double-peaked Balmer and He emission lines, and a
noticeable flux contribution from the low-mass companion star at the
longest wavelengths. In contrast, the BOSS spectrum is entirely
dominated by the two stellar components, with practically no evidence
for ongoing accretion~--~in fact, it resembles the plethora of
detached white dwarf/M-dwarf binaries identified by SDSS
\citep{silvestrietal06-1, rebassa-mansergasetal07-1,
  rebassa-mansergasetal10-1, rebassa-mansergasetal12-1}. The
normalised \Ha\ line profile in Fig.\,\ref{f-spectra} illustrates the
dramatic change in the strength and shape of the emission line.

In addition to the morphological change in the spectral appearance,
IR\,Com was fainter during the 2012 observations. The light curve of
IR\,Com obtained by the Catalina Real-Time Transient Survey (CRTS,
\citealt{drakeetal09-1}) between April 2005 and March 2013
(Fig.\,\ref{f-CRTS}) shows that the system was actively accreting
until $\sim$May 2011. From November 2011 to March 2013, IR\,Com was
persistently faint in 72 CRTS observations, which is to our knowledge
the longest, and best-recorded deep low state of this
system. Eliminating a few individual data points where IR\,Com was
unusually faint (which were taken during eclipse, see below), we find
an apparent low-state magnitude of $17.8$ in the unfiltered CRTS
photometry.

\section{Analysis}
\label{s-analysis}
\begin{figure}
\includegraphics[angle=-90,width=\columnwidth]{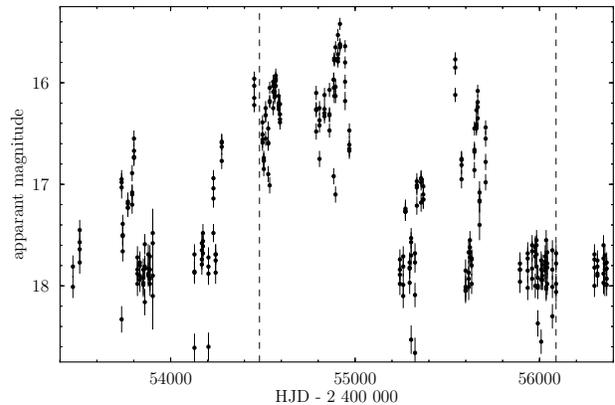}
\caption{\label{f-CRTS} CRTS light curve of IR\,Com, obtained between
  April 2005 and March 2013. The dashed lines mark the times at which
  SDSS spectra were taken (January 2008, and June 2012).}
\end{figure}

Several CRTS observations show IR\,Com fainter than the average
low-state magnitude, $\simeq17.8$ (Fig.\,\ref{f-CRTS}). Adopting the
ephemeris of \cite{felineetal05-1}, they all fall within the phase
interval of the eclipse, $\phi\simeq-0.04$ to $\phi\simeq+0.04$ (see
Fig.\,1 of \citealt{felineetal05-1}).

\begin{figure}
\includegraphics[angle=-90,width=\columnwidth]{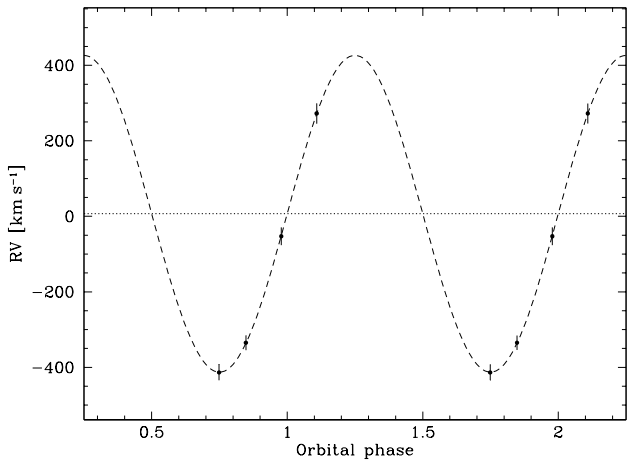}
\caption{\label{f-k2} The radial velocity of the M-dwarf in IR\,Com
  was measured using the \Lines{Na}{I}{8183.27,
    8194.81}\,\AA\ absorption doublet from three BOSS sub-spectra and
  one SDSS sub-spectrum (Table\,\ref{t-narv}). Fitting those
  velocities with a sinusoidal of fixed period results in a radial
  velocity amplitude of $K_2=419.6\pm3.4$\,km/s.}
\end{figure}

The \Lines{Na}{I}{8183.27, 8194.81}\,\AA\ absorption doublet of the
M-dwarf is detected in all three 2012 low-state SDSS sub-spectra, and
in one 2008 sub-spectrum. We fitted the \Ion{Na}{I} doublet with a
double-Gaussian profile of fixed separation (see
\citealt{rebassa-mansergasetal07-1} for details) to measure the radial
velocity variation of the companion star. These radial velocities
(Table\,\ref{t-narv}) were then fitted with a sine function, keeping
the period fixed to the value of \cite{felineetal05-1}. We find a
radial velocity amplitude of $K_2=419.6\pm3.4$\,km/s and a systemic
velocity of $\gamma=6.9\pm4.4$\,km/s. The phase of the radial velocity
curve agrees within 2\% with the expected blue-to-red crossing at
$\phi=0.0$ (Fig.\,\ref{f-k2}).

\begin{table}
\caption{\label{t-narv} Radial velocities of the companion star in
  IR\,Com measured from the \Lines{Na}{I}{8183.27,
    8194.81}\,\AA\ doublet. }
\begin{tabular}{lccc}
\hline
HJD & RV [km/s] & $\sigma$RV [km/s] \\
\hline
 2454480.995852 & -413.6 & 20.5 \\
 2456089.652378 & -334.8 & 19.3 \\
 2456089.663743 &  -52.8 & 23.6 \\
 2456089.675108 &  272.6 & 26.5 \\
\hline
\end{tabular}
\end{table}

The \Ha\ emission line in the bright state is strongly double-peaked
(inset in Fig.\,\ref{f-SDSS}), as expected for the origin from the
Keplerian motion in an accretion disc \citep{horne+marsh86-1}. During
the low state, the \Ha\ emission is much weaker and narrower, and its
radial velocity varies as expected for an origin on the inner
hemisphere of the companion star. However, there is clear evidence for
an asymmetry in the line profile, suggesting that emission from
another location within the system contributes. Such additional
\Ha\ components have been seen in a number of detached white dwarf /
M-dwarf binaries, and CVs in low states, where they originate either
on (or very close to) the white dwarf (e.g. \citealt{tappertetal07-1,
  tappertetal11-2, parsonsetal12-1, parsonsetal13-1}) or from material
located between the two stars (e.g. \citealt{gaensickeetal98-2,
  odonoghueetal03-1, kafkaetal05-2, parsonsetal11-1}).  Disentangling
the multiple \Ha\ components in IR\,Com, and identification of their
origin will require observations with higher spectral resolution than
the available SDSS low-state spectroscopy.

\section{Results and discussion}

\subsection{System parameters}
\label{s-syspar}

\begin{figure}
\includegraphics[angle=-90,width=\columnwidth]{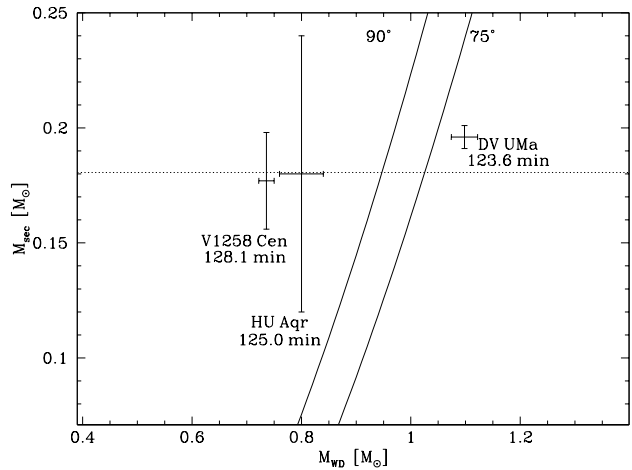}
\caption{\label{f-mfunc} The detection of white dwarf eclipses
  \citep{felineetal05-1} implies $i>75^{\circ}$. Together with the
  radial velocity amplitude of the companion star,
  $K_2=419.6\pm3.4$\,km/s, this limits the range of possible white
  dwarf masses. The dashed line is the donor mass for
  $P_\mathrm{orb}=2.09$\,h, according to the empirical CV evolution
  track of \citet{kniggeetal11-1}, and we show the component masses of
  three eclipsing CVs with very similar orbital periods to IR\,Com
  \citep{savouryetal11-1, schwopeetal11-1}.}
\end{figure}

The measured radial velocity amplitude of the companion star,
$K_2=419.6\pm3.4$\,km/s, can be used to compute the mass function of
the white dwarf. \cite{felineetal05-1} obtained high-time resolution
photometry of IR\,Com in quiescence, and showed that the morphology of
their light curve is consistent with a deep eclipse of the white dwarf
(as opposed to only the bright spot being eclipsed). This sets a
conservative limit on the binary inclination,
$75^{\circ}<i<90^{\circ}$, and the corresponding mass functions
(Fig.\,\ref{f-mfunc}) imply $0.8\,M_\odot<M_\mathrm{wd}<1.0\,M_\odot$,
which is fully consistent with the relatively high average CV white
dwarf mass found by \citet{zorotovicetal11-1} and
\citet{savouryetal11-1}. The companion mass can currently not be
constrained. For $P_\mathrm{orb}=2.09$\,h, the semi-empirical CV
evolution sequence of \citet{kniggeetal11-1} suggests
$M_2=0.186\,M_\odot$ (see Fig\,\ref{f-mfunc}), which would further
constrain the primary mass to
$0.95M_\odot<M_\mathrm{wd}<1.03\,M_\odot$.

\begin{figure}
\includegraphics[angle=-90,width=\columnwidth]{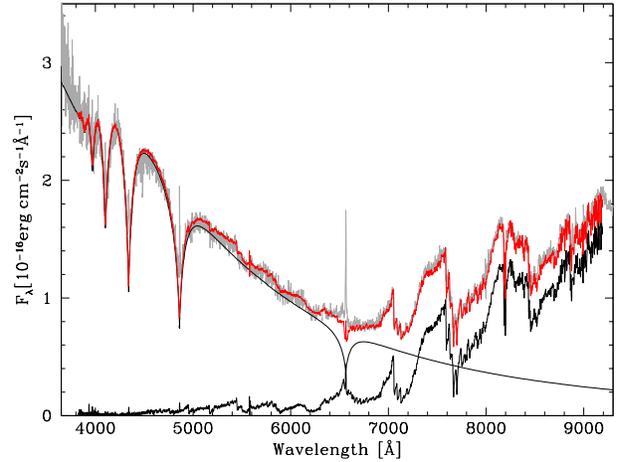}
\caption{\label{f-sfit} The low-state spectrum of IR\,Com (grey)
  obtained by SDSS can be modelled by the combination of a
  $\sim15000$\,K white dwarf and a M6--7 dwarf (both black, red is the
  sum of the two) at a distance of $d=115-165$\,pc. }
\end{figure}

We used the spectral decomposition method developed by
\citet{rebassa-mansergasetal07-1} for the analysis of detached white
dwarf/M-dwarf binaries to estimate the stellar parameters of IR\,Com
from the low-state SDSS spectrum (Fig.\,\ref{f-sfit}). In brief, the
composite SDSS spectrum is first fitted with two grids of white dwarf
and M-dwarf templates derived from SDSS spectroscopy, which gives the
spectral type and flux contribution of the companion. In a second
step, the best-fit M-dwarf template is subtracted from the composite
spectrum, and the residual spectrum is fitted with pure-hydrogen (DA)
white dwarf models from \citet{koester10-1}. We find that the
companion in IR\,Com has a spectral type of M6--7, which is later than
expected for its orbital period \citep{beuermannetal98-1,
  kniggeetal11-1}. The model atmosphere fit to the residual white
dwarf spectrum results in $T_\mathrm{wd}\simeq17\,500$\,K, and $\log
g=9.5$, which is the highest gravity in the model grid, corresponding
to a Chandrasekhar mass white dwarf. The white dwarf temperature, as
well as the implied secular accretion rate, are typical for the
orbital period range \citep{townsley+bildsten03-1,
  townsley+gaensicke09-1}. However, given that the mass function
implies $0.8\,M_\odot<M_\mathrm{wd}<1.0\,M_\odot$ it appears that the
fit to the Balmer lines substantially over-estimates the surface
gravity. One possibility is a magnetic field, too low to result in
visible Zeeman splitting in the Balmer lines, but sufficiently strong
to cause additional broadening of the lines, mimicking the higher
pressure in a more massive white dwarf. We return to this in
Sect.\,\ref{s-bfield}.  Another possibility is that He in the
atmosphere results in increased Stark broadening, as the
\Line{He}{I}{4471}\,\AA\ absorption line is visible in the BOSS
spectrum. A more definitive atmosphere model will require
better-quality low-state spectroscopy to fit also for the
He-abundance.

A crude estimate of the distance to IR\,Com can be estimated following
the prescription of \citet{beuermann06-1}, who made use of the fact
that the surface brightness near 7500\,\AA, and depth of the TiO band
near 7165\,\AA\ are a strong function of the spectral type, and
tabulated the relevant surface fluxes as a function of spectral
type. Thus measuring the mean observed fluxes in the bands
7450--7550\,\AA\ and 7140--7190\,\AA, and adopting
$R_2=0.213\,R_\odot$ (from Knigge's \citeyear{kniggeetal11-1}
sequence) gives $d=165$\,pc and 115\,pc for a companion spectral type
of M6 and M7, respectively (a spectral type of M5--M4, as typically
found near the lower edge of the period gap, would imply
$d=220-285$\,pc).

\subsection{A prolonged deep low state}
The long-term light curve of \citet{richteretal97-1}, spanning nearly
35\,years with rather sparse sampling, shows the system meandering
between 16\,mag and 17\,mag, with occasional drops to 18\,mag. So far,
only four short, bright states reaching 14\,mag have been reported (in
1959, 1988, 1996, and 2002, see \citealt{richteretal97-1} and
\citealt{katoetal02-2}), which led to the classification of IR\,Com as
a dwarf nova. However, while the eight years of CRTS data
(Fig.\,\ref{f-CRTS}) shows copious amounts of accretion activity, no
clearly defined dwarf nova-like outburst are obviously picked out,
confirming that the outburst recurrence time is very long.

The most puzzling feature in the CRTS light curve is the extremely
long low state, lasting over two years at the time of writing. Such
extended low states have so far been well-documented only among
polars \citep[e.g.][]{kafka+honeycutt05-1}, and among the VY\,Scl
stars, a sub-group of the nova-like variables
\citep[e.g.][]{honeycutt+kafka04-1}. The fundamental cause for the
occurrence of low states is not totally understood, but one hypothesis
is that star spots on the companion star forming at, or migrating into
the inner Lagrangian point lead to a sufficient depression of the
atmospheric scale height to reduce the mass loss, or stop it
all-together \citep{livio+pringle94-1, hessmanetal00-1}. This effect
should, in principle, be present in all CVs. In polars, the absence of
an accretion disc implies that a drop in the mass loss rate causes
also an almost immediate drop in the system's brightness. VY\,Scl
stars are normally in a state of high mass transfer, resulting in an
ionised, highly viscous, hot and bright accretion disc that will
rapidly drain onto the white dwarf once mass loss from the companion
ceases. The high temperature of the white dwarf
(e.g. \citealt{gaensickeetal99-1, araujo-betancoretal03-1,
  hoardetal04-1}) keeps any residual disc ionised, effectively
suppressing disc outbursts during the low state \citep{leachetal99-1}.

Low states in dwarf novae are extremely
rare. \citet{schreiberetal02-1} detected a low state in the
Z\,Cam-type dwarf nova RX\,And ($P_\mathrm{orb}=5.04$\,h, above the
period gap), lasting $\simeq200$\,d\footnote{Prior to the low state,
  RX\,And was in a stand-still, which may explain the low state
  analogous to the VY\,Scl stars discussed above, i.e. the rapid
  draining of a hot, highly viscous disc.}. Below the period gap,
there is no published evidence for well-defined, long-lasting low
states among dwarf novae, see the summary by \citet{warner99-1}. The
best-documented system is HT\,Cas, which underwent an outburst-free
period of $\sim4$\,years, however, still exhibiting $\sim1.5$\,mag
brightness fluctuations. The fundamental reason why dwarf novae,
despite most likely undergoing similar mass transfer variations as
polars or VY\,Scl stars, do not exhibit deep low states is that their
accretion discs buffer substantial amounts of mass, and only a small
fraction of mass is accreted onto the white dwarf during an individual
outburst (e.g. \citealt{cannizzo93-2}). \citet{schreiberetal00-1}
computed the outburst behaviour of a dwarf nova with a variable mass
transfer rate (empirically determined from the polar AM\,Her), and
found that outbursts never stop, even during prolonged states of zero
mass loss from the companion.

\subsection{A low magnetic field?}
\label{s-bfield}
Could the white dwarf in IR\,Com be magnetic?
\citet{katoetal02-2} argued against an IP nature of IR\,Com,
  based on the fact that the outburst they observed was substantially
  longer ($>4$\,d) compared to the very short ($\la1$\,d) outbursts
  seen among the handful of short-period IPs (e.g. EX\,Hya and
  HT\,Cam), and proposed IR\,Com to be a member of the SU\,UMa
  family\footnote{It is worth mentioning that no superhumps were
    detected during that outburst. If the true mass ratio of IR\,Com
    is close to the upper limit implied by our system parameters,
    $q<0.22$, (Sect.\,\ref{s-syspar}), it may lie above the threshold
    where tidal instabilities are triggered in the accretion disc
    \citep{whitehurst88-1}, even though superhumps have been detected
    in systems above the period gap having larger values of $q$
    \citep{pattersonetal05-3}.}  \citet{richteretal97-1} also argued
  against an IP nature of IR\,Com, based on the non-detection of a
  coherent (white dwarf spin) period in their optical data. One might
  take the CRTS light curve (Fig.\,\ref{f-CRTS}) as an additional
  argument against the IP nature of IR\,Com, as no deep long-lasting
  low states have been observed among IPs \citep{warner99-1}.

\begin{figure}
\includegraphics[angle=-90,width=\columnwidth]{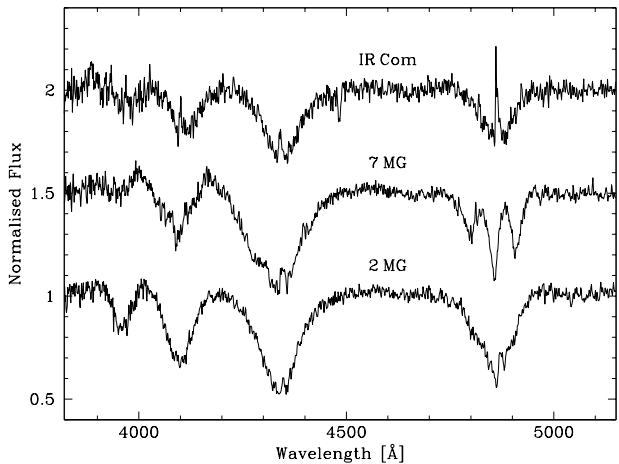}
\caption{\label{f-bfield} Normalised SDSS spectra of IR\,Com in the
  low state (top), and SDSS spectra of two DA white dwarfs from
  \citet{kawkaetal07-1} with comparable temperatures and field
  strengths of 7\,MG (SDSSJ124851.31-022924.7, middle), and 2\,MG
  (SDSSJ164703.21+370910.4, bottom). The absence of Zeeman splitting
  in the Balmer lines of IR\,Com suggests that the magnetic field
  strength of its white dwarf is $B\la5$\,MG.}
\end{figure}

The SDSS low-state spectrum of IR\,Com does not show any clear
evidence for Zeeman-splitting of the Balmer lines, and comparison to
the SDSS spectra of magnetic DA white dwarfs from
\citet{kawkaetal07-1} rules out a magnetic field in excess of
$\simeq5$\,MG (Fig.\,\ref{f-bfield}). This is less than the lowest
field detected among polars, $B\simeq7$\,MG in V2301\,Oph
\citep{ferrarioetal95-1}, but we can not rule out a field in the range
expected for IPs (a few 100\,kG to a few MG, see
\citealt{nortonetal04-1}).

\section{Conclusions}
We detected a deep, long low-state in the short-period CV
IR\,Com. Given that we can rule out a strong magnetic field on the
white dwarf, the occurrence of such a low state is so far unique among
the known population of short-period CVs, and we suspect that a weak
magnetic field on the white dwarf may be the cause for this unusual
behaviour. This possibility should be explored with higher-resolution
spectroscopy, or spectropolarimetry, capable of detecting sub-MG
fields. We also encourage additional observations during a low state
to refine the stellar masses and radii of the white dwarf and its low
mass companion.

Yet, (at least) one of the fundamental physical parameters in IR\,Com
must differ from those of ordinary SU\,UMa type dwarf novae below the
gap to explain its unusual long-term variability. We have shown that
the stellar masses and the accretion rate are normal for the orbital
period of the system, and hence, despite those (somewhat
circumstantial) arguments against IR\,Com being an IP, we believe that
a weak magnetic field on the white dwarf is the most plausible cause
for its observed behaviour

\section*{Acknowledgements}
CJM acknowledges an undergraduate research bursary from the Institute
of Physics. The research leading to these results has received funding
from the European Research Council under the European Union's Seventh
Framework Programme (FP/2007-2013) / ERC Grant Agreement n. 320964
(WDTracer).  BTG was supported in part by the UK’s Science and
Technology Facilities Council (ST/I001719/1).  Funding for SDSS-III
has been provided by the Alfred P. Sloan Foundation, the Participating
Institutions, the National Science Foundation, and the U.S. Department
of Energy Office of Science. The SDSS-III web site is
http://www.sdss3.org/. We thank Stuart Littlefair for a rapid and
constructive referee report.

\bibliographystyle{mn_new}
\bibliography{aamnem99,aabib,proceedings}

\bsp

\label{lastpage}

\end{document}